\documentclass[11pt,a4paper]{article}
\usepackage{jheppub}
\usepackage[T1]{fontenc}
\usepackage{bm}
\hypersetup{colorlinks=true,linkcolor=blue,citecolor=blue,urlcolor=blue}

\makeatletter
\global\def\@fpheader{\,}
\makeatother

\title{Galactic Endothermic Production and Exothermic Detection of Excited Dark Matter: Implications for LUX-ZEPLIN}

\author[a]{Chuan-Yang Xing}
\affiliation[a]{College of Science, China University of Petroleum (East China),
Qingdao 266580, China}
\emailAdd{cyxing@upc.edu.cn}

\abstract{%
The LUX-ZEPLIN experiment reported a candidate nuclear recoil at $248\,\mathrm{keV}$. This work proposes that endothermic self-scattering $\chi_1+\chi_1\to\chi_2+\chi_2$ in the Galaxy can produce an excited dark matter population, whose exothermic scattering $\chi_2+\mathrm{Xe}\to\chi_1+\mathrm{Xe}$ in xenon can account for the LZ candidate event. In a benchmark with a light scalar mediator, the conversion cross section is large enough for high-velocity Galactic dark matter particles to generate an excited fraction at the percent level. For dark matter masses near the TeV scale and mass splittings $\delta\gtrsim 500\,\mathrm{keV}$, the xenon recoil spectrum covers the candidate region and can yield an $\mathcal O(1)$ event count. The $\chi_2$ population produced in the Galaxy is colder than a relic population from the early Universe and yields a different nuclear recoil spectrum.
}

\begin{document}
\maketitle
\flushbottom

\section{Introduction}
\label{sec:introduction}

Astrophysical and cosmological observations require a dominant nonbaryonic dark matter (DM) component~\cite{Bauer:2017qwy, Profumo:2019ujg}.
Weakly interacting massive particles (WIMPs) are well-motivated DM candidates, as weak-scale masses and interactions can naturally yield the observed relic abundance through thermal freeze-out.
This possibility is being tested through searches for nuclear recoils in many low-background detectors~\cite{Cooley:2021rws}.

Recently, LUX-ZEPLIN (LZ) reported an event consistent with a nuclear recoil at $248\pm23_{\rm stat}\pm23_{\rm sys}\,\mathrm{keV}$~\cite{LZ:2026axp}.
A DM interpretation of this event should explain this high recoil without producing an excess at lower energies.
Conventional elastic scattering typically predicts a falling recoil spectrum and tends to overproduce lower-energy events.
This motivates mechanisms that reshape the recoil spectrum~\cite{Su:2026rwz,Freese:2026sga,Fan:2026kxx,Wu:2026nhi,DiMauro:2026ldr,Smirnov:2026aqk,McCabe:2026crm,Langhoff:2026ujr,Palmisano:2026kuj,Yin:2026jnn,Du:2026guj,Bisal:2026khf,Visinelli:2026kgt,Kotlarski:2026pep,Ahmed:2026qjg,Lee:2026xxh,Nomura:2026qyq,Wang:2026ytg,Bandyopadhyay:2026gjw,Borah:2026zwf,Lee:2026jxl,Okada:2026upm,Borah:2026ris,Lian:2026hpm,Yamashita:2026ump,Lee:2026wof,Okada:2026eol,Du:2026lpa,Zhu:2026dag,Das:2026uyy,Yuan:2026djt,Qi:2026vyp,Kumar:2026lgi,Cabo-Almeida:2026uqw,Barman:2026omh,Dent:2026bji,deLima:2026shq,Baer:2026fpy,Fan:2026hzw,He:2026idw,Aghaie:2026vsu,Uttayarat:2026isp,Das:2026buc,Lou:2026idn,Mahapatra:2026glu,Unwin:2026rdp,An:2026pkc,Elahi:2026vlm,Jeesun:2026vzo,Sheng:2026tqt,Chattopadhyay:2026ryw}.

A widely studied possibility is endothermic inelastic scattering, in which halo DM $\chi_1$ up-scatters to a heavier state $\chi_2$ in the detector~\cite{Su:2026rwz,Freese:2026sga,Fan:2026kxx,Wu:2026nhi,DiMauro:2026ldr,Smirnov:2026aqk,McCabe:2026crm,Langhoff:2026ujr,Palmisano:2026kuj,Yin:2026jnn,Du:2026guj,Bisal:2026khf,Visinelli:2026kgt,Kotlarski:2026pep,Ahmed:2026qjg,Lee:2026xxh,Nomura:2026qyq,Wang:2026ytg,Bandyopadhyay:2026gjw,Borah:2026zwf,Lee:2026jxl,Okada:2026upm,Borah:2026ris,Lian:2026hpm,Yamashita:2026ump,Lee:2026wof,Okada:2026eol,Du:2026lpa,Zhu:2026dag,Das:2026uyy,Yuan:2026djt,Qi:2026vyp,Kumar:2026lgi,Cabo-Almeida:2026uqw,Barman:2026omh,Das:2026buc,Mahapatra:2026glu,An:2026pkc}.
The mass splitting introduces a kinematic threshold that suppresses low-energy recoils.
Concrete realizations include Higgsino and other electroweak constructions~\cite{Freese:2026sga,Fan:2026kxx,Wu:2026nhi,DiMauro:2026ldr,Smirnov:2026aqk,Langhoff:2026ujr,Yin:2026jnn,Du:2026guj,Bisal:2026khf,Visinelli:2026kgt,Kotlarski:2026pep,Ahmed:2026qjg}, scalar and fermionic singlet--doublet sectors~\cite{Nomura:2026qyq,Wang:2026ytg,Bandyopadhyay:2026gjw,Borah:2026zwf,Lee:2026jxl,Okada:2026upm,Borah:2026ris,Lian:2026hpm}, and dark sectors with new vector or axionlike mediators~\cite{Su:2026rwz,Lee:2026xxh,Yamashita:2026ump,Lee:2026wof,Okada:2026eol,Du:2026lpa,Zhu:2026dag,Das:2026uyy,Yuan:2026djt,Qi:2026vyp,Kumar:2026lgi,Cabo-Almeida:2026uqw}.
This scenario can be tested by solar capture~\cite{Langhoff:2026ujr,Qi:2026vyp,Bose:2026ndd,Nguyen:2026lui}, annual modulation~\cite{McCabe:2026crm}, and the high-energy sideband~\cite{Dent:2026bji,Langhoff:2026ujr}.

Exothermic inelastic scattering offers a different explanation, in which an excited DM state $\chi_2$ down-scatters to a lighter state $\chi_1$~\cite{Dent:2026bji,deLima:2026shq,Baer:2026fpy,Fan:2026hzw,He:2026idw,An:2026pkc}.
The released mass splitting can produce high-energy recoils.
Its rate is proportional to the excited-state fraction.
The present-day excited population can originate from freeze-in at low reheating temperature~\cite{He:2026idw} or a residual abundance after conversion~\cite{Baer:2026fpy}.
Preserving this population requires sufficiently weak interactions between the two states.

However, small-scale halo structure motivates sizable DM self-interactions~\cite{Tulin:2017ara}.
If interactions of this strength are generic in the dark sector, conversion in the early Universe would be efficient, creating tension with a primordial origin of $\chi_2$.
We propose that endothermic self-scattering regenerates the excited population in the Galaxy, whose exothermic scattering from xenon nuclei can account for the LZ candidate event.
Galactic virial motion raises the kinetic energy of DM particles and opens the endothermic process $\chi_1+\chi_1\to\chi_2+\chi_2$.
This population produces exothermic nuclear recoils through $\chi_2+\mathrm{Xe}\to\chi_1+\mathrm{Xe}$.
In our setup, a light scalar mediates the self-interactions that produce $\chi_2$, while a vector portal mediates $\chi_2$ scattering in xenon.
We estimate Galactic production, infer the excited fraction, determine the $\chi_2$ velocity distribution, and calculate the recoil spectrum.
We note that Galactic production of excited dark matter through inelastic self-scattering has also been studied in connection with halo dynamics~\cite{Vogelsberger:2018bok,ONeil:2022sidm} and various indirect-detection signals, including the 511-keV line~\cite{Finkbeiner:2007kk,Chen:2009dm,Chen:2009av,Morris:2011dj}, the 3.5-keV X-ray line~\cite{Finkbeiner:2014sja,Cline:2014kaa}, and coannihilation signals~\cite{Berlin:2023qco,Berlin:2025fwx}.

\section{Kinematics}
\label{sec:kinematics}

The dark sector contains a DM ground state $\chi_1$ and a heavier excited state $\chi_2$.
Their masses are separated by a positive splitting, $m_2=m_1+\delta$ with $\delta>0$.
Galactic production and xenon detection are connected by the sequence of inelastic reactions
\begin{equation}
 \chi_1+\chi_1\longrightarrow\chi_2+\chi_2,
 \qquad
 \chi_2+\mathrm{Xe}\longrightarrow\chi_1+\mathrm{Xe},
 \label{eq:two-step-processes}
\end{equation}
where $\mathrm{Xe}$ denotes a xenon nucleus.
The first reaction is endothermic because two mass splittings must be supplied by the relative motion of the incoming particles.
The second reaction is exothermic and releases one mass splitting into the kinetic energies of the outgoing DM particle and the recoiling nucleus.
For the TeV-scale DM and mass splitting $\delta\gtrsim 500\,\mathrm{keV}$ considered below, both steps are accurately described by nonrelativistic kinematics.

For the production process of a $\chi_2$ pair, let $u\equiv|\bm v_a-\bm v_b|$ denote the relative speed of the two incident $\chi_1$ particles.
Their reduced mass is $\mu_{11}=m_1/2$, and the center-of-mass kinetic energy in the relative motion is $E_{\rm rel}=m_1u^2/4$.
Opening the $\chi_2\chi_2$ channel requires $E_{\rm rel}\geq2\delta$.
The corresponding threshold is therefore
\begin{equation}
 u\geq u_{\rm th},
 \qquad
 u_{\rm th}=\sqrt{\frac{8\delta}{m_1}}.
 \label{eq:upscattering-threshold}
\end{equation}
For $m_1=4\,\mathrm{TeV}$ and $\delta=500\,\mathrm{keV}$, this gives $u_{\rm th}=299.8\,\mathrm{km/s}$.
The threshold lies within the range of Galactic relative velocities but selects the faster part of the incident $\chi_1$ population.

The second process in Eq.~\eqref{eq:two-step-processes} describes an incident $\chi_2$ with laboratory speed $v$ scattering from a stationary nucleus of mass $m_A$.
Energy and momentum conservation imply that at a fixed incident speed the recoil energy is restricted to $E_R^-(v)\leq E_R\leq E_R^+(v)$, with
\begin{equation}
 E_R^\pm(v)=
 \frac{\mu_{2A}^2v^2}{2m_A}
 \left(
  1\pm\sqrt{1+\frac{2\delta}{\mu_{2A}v^2}}
 \right)^2.
 \label{eq:exothermic-recoil-endpoints}
\end{equation}
Here $\mu_{2A}=m_2m_A/(m_2+m_A)$ is the reduced mass of the $\chi_2$--nucleus system.
The two endpoints correspond to the limiting orientations of the momentum transfer relative to the incident momentum.
As the incident speed decreases, the interval between them narrows and both endpoints approach $E_R^\star=(\mu_{2A}/m_A)\delta$.
This is the recoil energy obtained in the zero-velocity limit.
Finite incident velocities broaden the allowed recoil interval around this characteristic value.
This allowed recoil window permits an exothermic-scattering interpretation of the LZ candidate event when $E_R^-(v)\leq248\,\mathrm{keV}\leq E_R^+(v)$.

In this work, the relatively large mass splitting renders the inverse endothermic process $\chi_1+\mathrm{Xe}\to\chi_2+\mathrm{Xe}$ kinematically inaccessible, as its threshold speed $v_{\rm th}^{\rm endo}=\sqrt{2\delta/\mu_{1A}}$, where $\mu_{1A}=m_1m_A/(m_1+m_A)$, exceeds the maximum halo-DM speed.
For $m_1\gg m_A$, one has $v_{\rm th}^{\rm endo}/u_{\rm th}\simeq\sqrt{m_1/(4m_A)}\gg1$, with $u_{\rm th}$ given in Eq.~\eqref{eq:upscattering-threshold}, so the $\chi_1+\chi_1\to\chi_2+\chi_2$ channel can be open while the $\chi_1+\mathrm{Xe}\to\chi_2+\mathrm{Xe}$ channel remains closed.

\section{Galactic production and abundance of \texorpdfstring{$\chi_2$}{chi2}}
\label{sec:production}

Producing an appreciable Galactic $\chi_2$ abundance requires a sufficiently strong $\chi_1+\chi_1\to\chi_2+\chi_2$ interaction.
As a benchmark realization, we take this process to be mediated by a CP-even real scalar $\phi$.
The interaction Lagrangian relevant to this production process is
\begin{equation}
 \mathcal L_{\phi}=
 -\frac12g_{11}\phi\bar\chi_1\chi_1
 -\frac12g_{22}\phi\bar\chi_2\chi_2
 -g_{12}\phi\bar\chi_1\chi_2 .
 \label{eq:scalar-production-lagrangian}
\end{equation}
These interactions allow transitions from the $\lvert\chi_1\chi_1\rangle$ channel to both the mixed channel $\lvert\chi_1\chi_2\rangle_S\equiv(\lvert\chi_1\chi_2\rangle+\lvert\chi_2\chi_1\rangle)/\sqrt{2}$ and $\lvert\chi_2\chi_2\rangle$.
We take the couplings to be real and assume $g_{11}\ll1$, which suppresses the direct $\lvert\chi_1\chi_1\rangle\to\lvert\chi_1\chi_2\rangle_S$ transition.
Since the excited population is small, we also neglect reactions initiated by $\lvert\chi_2\chi_2\rangle$.
As a benchmark, we omit the mixed channel $\lvert\chi_1\chi_2\rangle_S$ and project the two-body dynamics onto the $\{\lvert\chi_1\chi_1\rangle,\lvert\chi_2\chi_2\rangle\}$ subspace.

Because $m_\phi\ll m_1$ and Galactic DM is nonrelativistic, scalar exchange is described by a Yukawa potential with range $m_\phi^{-1}$.
Moreover, the benchmark lies in the nonperturbative regime $\alpha_{22}\mu_{11}/m_\phi\gg1$~\cite{Tulin:2013teo, Schutz:2014nka}, where repeated scalar exchange cannot be captured by a tree-level Born cross section.
The projected two-channel wavefunction therefore obeys the coupled Schr\"odinger equation 
\begin{equation}
 \left[
 -\frac{\nabla^2}{2\mu_{11}}\bm 1+U_2(r)
 \right]\bm\Psi(\bm r)
 =E_{\rm rel}\bm\Psi(\bm r),
 \label{eq:coupled-schrodinger}
\end{equation}
where $r$ is the separation between the two DM particles, while the reduced mass $\mu_{11}$ and relative kinetic energy $E_{\rm rel}$ were defined in section~\ref{sec:kinematics}.
The potential matrix entering Eq.~\eqref{eq:coupled-schrodinger} is
\begin{equation}
 U_2(r)=
 \begin{pmatrix}
  0&W(r)\\
  W(r)&\Delta(r)
 \end{pmatrix},
 \qquad
 W(r)=-\alpha_{12}\frac{e^{-m_\phi r}}{r},
 \qquad
 \Delta(r)=2\delta-\alpha_{22}\frac{e^{-m_\phi r}}{r},
 \label{eq:projected-potential}
\end{equation}
where $\alpha_{ij}=g_{ij}^2/(4\pi)$.

Since $U_2(r)$ depends on the interparticle separation, its eigenvalues and eigenstates vary at different $r$.
At large separations, $W(r)\to0$ and $\Delta(r)\to2\delta$, so the eigenstates approach the diabatic channel states $\lvert\chi_1\chi_1\rangle$ and $\lvert\chi_2\chi_2\rangle$, with energies $0$ and $2\delta$, respectively.
The incoming particles therefore occupy the lower $\lvert\chi_1\chi_1\rangle$ channel.
As the particles approach each other, the attractive diagonal potential lowers the diabatic energy of the $\lvert\chi_2\chi_2\rangle$ channel until it becomes degenerate with the $\lvert\chi_1\chi_1\rangle$ channel at $\Delta(r_c)=0$, where $r_c$ is the crossing radius.
The off-diagonal coupling $W(r_c)$ turns this level crossing into an avoided crossing.
This avoided crossing is characterized by a minimum eigenvalue separation $2|W_c|$ at $r_c$, where the corresponding eigenstates are equal superpositions of the two diabatic channel states.
The minimum eigenvalue separation is determined by the off-diagonal coupling evaluated at the crossing radius:
\begin{equation}
 |W_c| = |W(r_c)|=2\delta\frac{\alpha_{12}}{\alpha_{22}}.
 \label{eq:crossing-quantities}
\end{equation}

Following the semiclassical classical-path treatment of nonadiabatic collisions~\cite{Nakamura:2012nonadiabatic}, we prescribe the relative coordinate by a classical trajectory while evolving the two channel amplitudes quantum mechanically.
For an encounter with relative speed $u$ and impact parameter $b$, the straight-line approximation gives $r(t)=\sqrt{b^2+u^2t^2}$.
A trajectory with $b<r_c$ crosses the avoided crossing twice, once on the incoming passage and once on the outgoing passage.
Near $r_c$, the diabatic level separation varies with the local slope $\Delta_c'\equiv(d\Delta/dr)_{r_c}=2\delta(m_\phi+1/r_c)$.
Together with the radial crossing speed, this slope determines the single-passage Landau--Zener probability of remaining in the same diabatic channel~\cite{Landau:1932,Zener:1932,Nakamura:2012nonadiabatic}:
\begin{equation}
 p(b,u)=e^{-2\pi\gamma(b,u)}.
 \label{eq:single-passage-probability}
\end{equation}
Here $\gamma(b,u)=|W_c|^2/(\Delta_c'|\dot r|_c)$ is the Landau--Zener parameter, with $|\dot r|_c=u\sqrt{1-b^2/r_c^2}$ denoting the radial speed evaluated at $r=r_c$ for the straight-line trajectory.
The two passages provide two histories leading to a final $\lvert\chi_2\chi_2\rangle$ state: the system can remain in the same diabatic channel at the first crossing and change channel at the second, or change at the first and remain at the second.
These histories interfere for an individual trajectory, but the broad distributions of velocities, impact parameters, production positions, and orbital histories wash out their relative phase in the Galactic ensemble.
After averaging over the St\"uckelberg phase, the double-passage conversion probability becomes~\cite{Nakamura:2012nonadiabatic}
\begin{equation}
 \overline P_{11\to22}(b,u)
 =2p(b,u)\bigl[1-p(b,u)\bigr].
 \label{eq:phase-averaged-conversion}
\end{equation}
In the strongly diabatic limit, $p\to1$ and the system remains in the initial channel at both crossings.
In the strongly adiabatic limit, $p\to0$ and the system changes diabatic channel at both crossings, returning to the initial channel after the second passage.
The net conversion probability is consequently largest at intermediate coupling and reaches $\overline P_{11\to22}=1/2$ for $p=1/2$.
Since $\gamma\propto|W_c|^2\propto\alpha_{12}^2$, this result predicts a nonmonotonic dependence of the $\chi_2$ production rate on $\alpha_{12}$.

Integrating Eq.~\eqref{eq:phase-averaged-conversion} over the impact parameter gives the production cross section
\begin{equation}
 \sigma_{11\to22}(u)=
 4\pi\Theta(u-u_{\rm th})
 \int_0^{r_c}b\,db\,
 p(b,u)\bigl[1-p(b,u)\bigr],
 \label{eq:production-cross-section}
\end{equation}
where the threshold speed $u_{\rm th}$ is given in Eq.~\eqref{eq:upscattering-threshold}.
The scale of the cross section is set by the geometric area $\pi r_c^2$.
As a numerical benchmark, taking $m_\phi=1\,\mathrm{MeV}$, $\delta=500\,\mathrm{keV}$, and $\alpha_{22}=0.1$ gives $r_c\simeq18.0\,\mathrm{fm}$, corresponding to a characteristic geometric scale $\pi r_c^2\simeq 10^{-23}\,\mathrm{cm^2}$ for the production cross section.
If the outgoing scattering angle is tracked for each impact parameter, the differential cross section can also be obtained.

We assume a spatially homogeneous density, an initially negligible excited population, and a $\chi_2$ lifetime longer than the Galactic production time. We also assume only a small fraction of $\chi_1$ converts, and its number density $n_1$ is approximately constant.
The characteristic upscattering rate per particle is
\begin{equation}
 \Gamma_{\rm up}
 =n_1\left\langle\sigma u\right\rangle_{11\to22},
 \label{eq:upscattering-rate}
\end{equation}
The velocity average entering Eq.~\eqref{eq:upscattering-rate} is defined by
\begin{equation}
 \left\langle\sigma u\right\rangle_{11\to22}
 \equiv
 \frac{1}{n_1^2}\int\mathrm d^3v_a\,\mathrm d^3v_b\,
 F_1(\bm v_a)F_1(\bm v_b)
 \sigma_{11\to22}(u)u,
 \label{eq:velocity-averaged-production}
\end{equation}
Here $F_1(\bm v)$ is the velocity distribution of $\chi_1$, with $\int\mathrm d^3v\,F_1(\bm v)=n_1$.
The endothermic threshold therefore makes the abundance sensitive to the high-relative-speed part of the Galactic distribution, especially for larger splittings.

\begin{figure}[!t]
 \centering
 \includegraphics[width=0.75\textwidth]{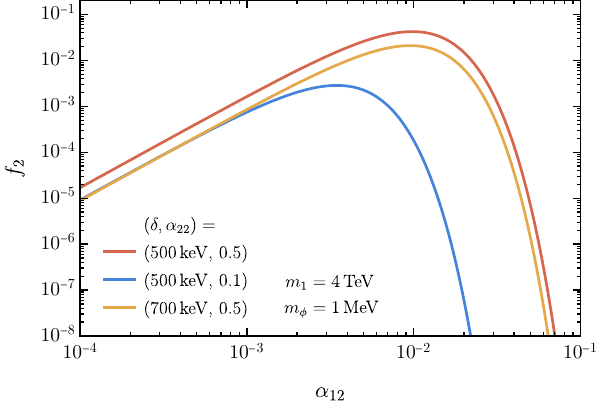}
 \caption{Fraction $f_2$ of Galactic DM in the excited state $\chi_2$ as a function of $\alpha_{12}$ in the projected two-channel benchmark for $m_1=4\,\mathrm{TeV}$, $m_\phi=1\,\mathrm{MeV}$, and local DM density $\rho_\chi=0.3\,\mathrm{GeV/cm^3}$. The nonmonotonic behavior results from the suppression of the double-passage probability in both the diabatic and adiabatic limits.}
 \label{fig:chi2-fraction-alpha12}
\end{figure}

The accumulated excited-state fraction is estimated as
\begin{equation}
 f_2\equiv\frac{n_2}{n_1+n_2}
 \simeq\Gamma_{\rm up}t_{\rm MW},
 \label{eq:excited-fraction-estimate}
\end{equation}
Here $n_2$ is the produced $\chi_2$ number density, and $t_{\rm MW}$ denotes the effective duration over which Galactic production accumulates, for which we adopt the representative value $t_{\rm MW}\simeq10\,\mathrm{Gyr}$.
This approximation applies for $\Gamma_{\rm up}t_{\rm MW}\ll1$.

Figure~\ref{fig:chi2-fraction-alpha12} quantifies the predicted nonmonotonic dependence of the excited fraction on the off-diagonal coupling $\alpha_{12}$ for three choices of $\delta$ and $\alpha_{22}$.
For $\alpha_{22}=0.5$, the maximum fraction decreases from approximately $4.2\%$ at $\delta=500\,\mathrm{keV}$ to $2.1\%$ at $\delta=700\,\mathrm{keV}$ because the larger splitting raises the threshold relative speed for production.
Reducing $\alpha_{22}$ to $0.1$ decreases the crossing radius and lowers the maximum fraction to approximately $0.29\%$.

\section{Velocity distribution of \texorpdfstring{$\chi_2$}{chi2}}
\label{sec:chi2-distribution}

The estimate above gives the excited fraction but not the $\chi_2$ velocity distribution at the Solar position.
These velocities are first determined by the conversion event: only sufficiently fast $\chi_1$ pairs can overcome the threshold, and part of their relative kinetic energy becomes excitation energy.
The newly produced $\chi_2$ particles then travel through the Galactic potential, which further changes their velocities.

For a given $\chi_2$ velocity at the Solar position, we trace the corresponding orbit backward.
The present $\chi_2$ distribution $F_2$ is obtained by accumulating production along that orbit:
\begin{equation}
 F_2(\bm x_\odot,\bm v,t_0)
 =\int_{t_0-t_{\rm MW}}^{t_0}dt\,Q_2(\bm x(t),\bm v(t)),
 \label{eq:chi2-characteristic}
\end{equation}
Here $(\bm x(t),\bm v(t))$ is the orbit satisfying $\bm x(t_0)=\bm x_\odot$ and $\bm v(t_0)=\bm v$, where $\bm x_\odot$ is the Solar position and $t_0$ is the present time.
The source $Q_2$, the $\chi_2$ production rate per unit phase-space volume, sums over incident $\chi_1$ pairs and outgoing directions:
\begin{equation}
 \begin{split}
 Q_2(\bm x,\bm w)={}&\frac12\int d^3v_a\,d^3v_b\,
 F_1(\bm x,\bm v_a)F_1(\bm x,\bm v_b)\,u\\
 &\times\int d\Omega_{\rm sc}\,
 \frac{d\sigma_{11\to22}}{d\Omega_{\rm sc}}
 \sum_{\pm}\delta^{(3)}(\bm w-\bm v_2^\pm).
 \end{split}
 \label{eq:chi2-source}
\end{equation}
Here $\bm x$ is the production position, $\bm w$ is the velocity of a produced $\chi_2$, and $F_1$ is the $\chi_1$ number density in phase space.
The incident velocities $\bm v_a$ and $\bm v_b$ have relative speed $u=|\bm v_a-\bm v_b|$.
The solid angle $\Omega_{\rm sc}$ specifies the direction of one outgoing $\chi_2$ in the pair's center-of-mass frame.
The sum includes both outgoing particles with velocities $\bm v_2^\pm$, while $1/2$ counts each incident pair once.

\begin{figure}[t]
 \centering
 \includegraphics[width=0.75\textwidth]{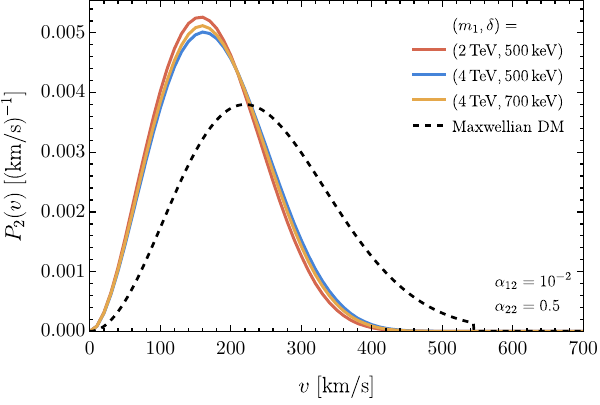}
 \caption{Normalized $\chi_2$ speed distributions at the Solar position for the three indicated $(m_1,\delta)$ points, with $(\alpha_{12},\alpha_{22})=(0.01,0.5)$. The dashed curve is a separately normalized truncated Maxwellian reference with $v_0=220\,\mathrm{km/s}$ and $v_{\rm esc}=544\,\mathrm{km/s}$.}
 \label{fig:chi2-speed}
\end{figure}

The source $Q_2$ requires a Galactic halo model for $\chi_1$.
We adopt an NFW number-density profile and obtain the velocity distribution by isotropic Eddington inversion in the static spherical Galactic potential of Ref.~\cite{Ferrer:2013cla}.
The same potential determines the $\chi_2$ orbits.
Since the excited fraction is small, we hold $F_1$ fixed, so $Q_2$ is time independent.

At the Solar position, the normalized speed distribution is
\begin{equation}
 P_2(v)=\frac{v^2}{n_2(\bm x_\odot)}
 \int d\Omega_{\bm v}\,F_2(\bm x_\odot,\bm v,t_0),
 \label{eq:chi2-speed-distribution}
\end{equation}
Here $n_2(\bm x_\odot)$ is the local $\chi_2$ number density, $v=|\bm v|$, and the integral runs over directions of $\bm v$.
Figure~\ref{fig:chi2-speed} shows the $\chi_2$ speed distributions for three benchmark points.
They are colder than the conventional truncated Maxwellian DM distribution because excitation removes energy from the incident relative motion.
The normalized $\chi_2$ curves also differ little across the three benchmarks.
Increasing $\delta/m_1$ raises the threshold, excluding more incident pairs.
The pairs that can still scatter have higher relative speeds, so the difference $u^2-u_{\rm th}^2$, which sets the outgoing relative speed, changes only modestly.

\section{Xenon recoil spectrum}
\label{sec:recoil}

The local $\chi_2$ population obtained in section~\ref{sec:chi2-distribution} can be probed through the exothermic scattering process in Eq.~\eqref{eq:two-step-processes}.
For the Majorana states $\chi_1$ and $\chi_2$, the diagonal vector currents vanish identically, whereas the off-diagonal transition current remains allowed.
Their interaction with nucleons through a vector mediator is
\begin{equation}
 \mathcal L_{V}
 =i g_\chi V_\mu\bar\chi_2\gamma^\mu\chi_1
 +V_\mu\left(
  g_p\bar p\gamma^\mu p
  +g_n\bar n\gamma^\mu n
 \right).
 \label{eq:vector-portal}
\end{equation}
Here $V_\mu$ is the vector mediator with mass $m_V$, $g_\chi$ is the transition coupling, and $g_p$ and $g_n$ are its couplings to protons and neutrons.
For a nucleus with atomic number $Z$ and mass number $A$, the coherent coupling is $g_A=Zg_p+(A-Z)g_n$.
We specialize to a dark photon portal, for which $g_p=\epsilon e$, $g_n\simeq0$, and hence $g_A=Z\epsilon e$.

The differential cross section for scattering from a xenon isotope is
\begin{equation}
 \frac{\mathrm d\sigma_{2A\to1A}}{\mathrm dE_R}
 =\frac{m_A}{2\pi v^2}
 \frac{g_\chi^2g_A^2}
 {\left(m_V^2+2m_AE_R\right)^2}
 F_A^2(q)\,
 \Theta\!\left[v-v_{\min}^{\rm exo}(E_R)\right],
 \label{eq:exothermic-cross-section}
\end{equation}
where $q^2=2m_AE_R$ and $F_A(q)$ is the nuclear form factor.
Energy and momentum conservation give the minimum incident speed compatible with a recoil energy $E_R$,
\begin{equation}
 v_{\min}^{\rm exo}(E_R)
 =\frac{\left|m_AE_R/\mu_{2A}-\delta\right|}
 {\sqrt{2m_AE_R}},
 \label{eq:exothermic-vmin}
\end{equation}

The recoil spectrum per unit detector mass follows by folding Eq.~\eqref{eq:exothermic-cross-section} with the laboratory-frame $\chi_2$ velocity distribution $F_{2,\mathrm{lab}}(\bm v)$,
\begin{equation}
 \frac{\mathrm dR}{\mathrm dE_R}
 =\sum_A\frac{\xi_A}{m_A}
 \int_{v>v_{\min}^{\rm exo}(E_R)}\!\mathrm d^3v\,
 F_{2,\mathrm{lab}}(\bm v)\,v\,
 \frac{\mathrm d\sigma_{2A\to1A}}{\mathrm dE_R},
 \label{eq:recoil-rate}
\end{equation}
where $\xi_A$ is the mass fraction of isotope $A$.
Eq.~\eqref{eq:recoil-rate} is evaluated by summing over the xenon isotopes and using the standard Helm form factor.
The isotope dependence is important near the candidate recoil energy, where $q\simeq0.246\,\mathrm{GeV}$ and finite nuclear size strongly modifies the spectrum.
The resulting shape is determined jointly by the exothermic minimum speed, the incident $\chi_2$ distribution, the mediator propagator, and the xenon nuclear response.

\begin{figure}[t]
 \centering
 \includegraphics[width=0.75\textwidth]{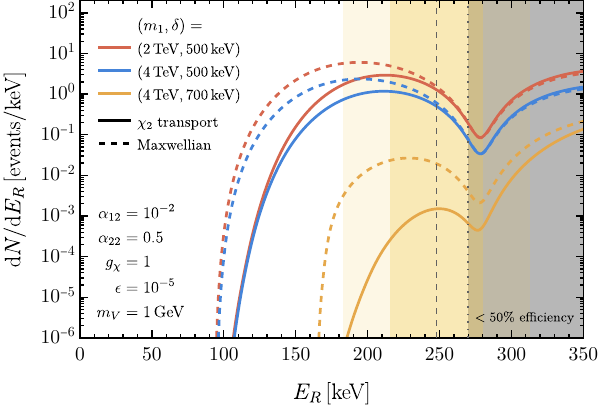}
 \caption{Xenon recoil spectra $dN/dE_R$ for three $(m_1,\delta)$ points. Solid curves use the transported $\chi_2$ distributions and their calculated local densities; dashed curves use a truncated Maxwellian with the same density for each point. The vertical dashed line marks the $248\,\mathrm{keV}$ LZ candidate, the yellow bands its $1\sigma$ and $2\sigma$ intervals, and the dotted line and gray region the nominal 50\% efficiency boundary.}
 \label{fig:recoil-spectrum}
\end{figure}

For an exposure $\mathcal E$, the recoil event spectrum is
\begin{equation}
 \frac{\mathrm dN}{\mathrm dE_R}
 =\mathcal E\frac{\mathrm dR}{\mathrm dE_R}.
 \label{eq:event-spectrum}
\end{equation}
Figure~\ref{fig:recoil-spectrum} compares recoil spectra for three benchmark choices of $(m_1,\delta)$.
For this comparison, the local excited-state density is calculated for each benchmark.
The portal parameters $g_\chi=1$, $\epsilon=10^{-5}$, and $m_V=1\,\mathrm{GeV}$ are consistent with existing constraints~\cite{Fabbrichesi:2020wbt}.
The LZ exposure of $2.84\,\mathrm{tonne\,years}$ is adopted~\cite{LZ:2026axp}.
The benchmark spectra have support across the LZ candidate region, as allowed by the exothermic recoil kinematics discussed in section~\ref{sec:kinematics}.
At lower recoil energies, the spectra are kinematically cut off.
The spectra also extend beyond the nominal 50\% efficiency boundary, but the rapidly decreasing detector acceptance strongly suppresses the visibility of these higher-energy recoils.
The spectrum normalization scales as $g_\chi^2\epsilon^2$, so suitable portal couplings can yield an $\mathcal O(1)$ expected event count in the LZ candidate region.

The dashed curves use a truncated Maxwellian distribution as a reference for $\chi_2$ produced in the early Universe~\cite{deLima:2026shq,Baer:2026fpy}.
Since Galactic production yields a colder $\chi_2$ population, the lower recoil-energy edge of the solid curves lies at higher energy.
This difference in spectral shape could distinguish the two origins of excited DM.

We calculate the lifetime of $\chi_2$ and verify that it exceeds the age of the Milky Way.
Since $\delta<m_\phi,m_V,2m_e$, the two-body decays $\chi_2\to\chi_1+\phi$ and $\chi_2\to\chi_1+V$, as well as $\chi_2\to\chi_1+e^++e^-$, are kinematically closed. Kinetic mixing also couples $V$ to electrons, allowing the loop-induced decay $\chi_2\to\chi_1+3\gamma$, with leading-order decay width~\cite{Krnjaic:2025zjl},
\begin{equation}
 \Gamma_{\chi_2}
 \simeq\frac{1}{6.8\times10^{4}\,\mathrm{Gyr}}
 \left(\frac{\epsilon g_\chi}{10^{-5}}\right)^2
 \left(\frac{\delta}{500\,\mathrm{keV}}\right)^{13}
 \left(\frac{1\,\mathrm{GeV}}{m_V}\right)^4.
 \label{eq:chi2-lifetime}
\end{equation}
Thus, $\chi_2$ is sufficiently long-lived over the Galactic timescale $t_{\rm MW}\simeq 10\,\mathrm{Gyr}$.

\section{Conclusions}
\label{sec:conclusions}

This work presents a two-state DM scenario in which endothermic DM self-scattering in the Galaxy produces an excited-state population, followed by vector-mediated exothermic scattering in xenon detectors. For a benchmark with a light scalar mediator, the conversion cross section is large, and the Galactic high-velocity tail provides sufficient kinetic energy to overcome the excitation threshold, yielding a percent-level excited-state abundance. For TeV-scale DM and mass splittings $\delta\gtrsim 500\,\mathrm{keV}$, the resulting recoil spectra cover the $248\,\mathrm{keV}$ LZ candidate region, with a low-energy kinematic cutoff.

The present analysis uses a semiclassical two-channel description to estimate the production cross section. A full three-channel treatment would refine the predicted local abundance and recoil spectrum.

\acknowledgments
This work is supported by the Fundamental Research Funds for the Central Universities (No. 24CX06048A) and the Taishan Scholar Young Talent Program (Grant No. tsqn202507117).

\bibliographystyle{JHEP}
\bibliography{references}

\end{document}